\documentclass[aps,twocolumn,showpacs,nofootinbib,superscriptaddress]{revtex4}
\usepackage{graphicx}
\usepackage{amsmath}
\usepackage{amsfonts}
\usepackage{amssymb}
\usepackage{hyperref}

\usepackage{multirow}

\newcommand{\ketbra}[2]{\left| {#1} \right\rangle\left\langle {#2}
\right|}

\newcommand{\beq}{\begin{equation}}
\newcommand{\eeq}{\end{equation}}
\newcommand{\bqa}{\begin{eqnarray}}
\newcommand{\eqa}{\end{eqnarray}}

\begin{document}

\title{Extrema of discrete Wigner functions and applications}

\author{Andrea Casaccino}
\affiliation {Information Engineering Department, University of
Siena \\Via Roma 56, 53100 Siena, Italy}
\author{Ernesto F. Galv\~{a}o}
\affiliation{Instituto de F{\'i}sica, Universidade Federal
Fluminense \\Av. Gal. Milton Tavares de Souza s/n\\Gragoat{\'a},
Niter{\'o}i, RJ, 24210-340, Brazil}
\author{Simone Severini}
\affiliation{Institute for Quantum
Computing and Department of Combinatorics and Optimization,
University of Waterloo \\ 200 University Ave. W. N2L 3G1, Waterloo, Canada}

\date{\today}

\begin{abstract}
We study the class of discrete Wigner functions proposed by Gibbons \textit{et al.} [Phys. Rev. A 70, 062101 (2004)] to describe quantum states using a discrete phase-space based on finite fields. We find the extrema of such functions for small Hilbert space dimensions, and present a quantum information application: a construction of quantum random access codes. These are constructed using the complete set of phase-space point operators to find encoding states and to obtain the codes' average success rates for Hilbert space dimensions $2,3,4,5,7$ and $8$.
\end{abstract}
\maketitle

\section{Introduction}

The Wigner function $W(q,p)$ was introduced by Wigner in 1932 \cite{Wigner32} as a way to represent quantum states of one or more particles in phase-space. It is a quasi-probability distribution, which means it retains some of the properties of a true probability distribution, while having some surprising properties due to quantum effects. For example, it can be negative in some regions in phase-space.

There were many proposals of analogues of $W(q,p)$ to represent quantum systems with discrete degrees of freedom such as spins (see \cite {GibbonsHW04} and references therein for a review). These discrete Wigner functions have been applied to visualize quantum states and operations in the context of quantum information and computation \cite{BianucciMPS02} \cite{MiquelPS02} \cite{PazRS05}.

In this paper we study properties and applications of the class of discrete Wigner functions defined by Gibbons \textit{et al.} \cite{GibbonsHW04}, which take values on a discrete phase-space built with finite fields. We start in sections \ref{sec def} and \ref{sec mub} by reviewing the definition of this class of functions. In section \ref{sec spectra} we calculate the spectra of the phase-space point operators used to define the discrete Wigner functions. In section \ref{sec extre} we describe how to use the spectra calculated to find the extremal values for the discrete Wigner function for some small Hilbert space dimensions. We also describe how the phase-space point operators can be used in a quantum information application known as quantum random access codes. In section \ref{sec qrac} we introduce these codes with a simple example, and present a quantum random access code construction based on states that maximize the discrete Wigner function.

\section{Defining a class of discrete Wigner functions}\label{sec def}

The discrete phase-space is a $d \times d$ grid in which we identify some particular sets of $d$ points called \textit{lines}. \textit{Parallel lines} are lines sharing no points in common. Following Gibbons \textit{et al.} \cite{GibbonsHW04}, a partition of the $d^2$ phase-space points into $d$ parallel lines of $d$ points each will be called a \textit{striation}. The definition of lines and striations is done in such a way as to ensure, in this discrete geometry, some geometrical properties akin to the properties of lines in usual geometry:

\begin{description}
\item[i] given any two points, exactly one line contains both points;\label{pstria1}
\item[ii] given a point $\alpha$ and a line $\lambda$ not containing $\alpha$, there is exactly one line parallel to $\lambda$ that contains $\alpha$;\label{pstria2}
\item[iii] two non-parallel lines intersect at exactly one point. \label{pstria3}
\end{description}

Gibbons \textit{et al.} described how to define $d(d+1)$ lines, partitioned into $d+1$ striations of $d$ lines each, satisfying the requirements above. The construction is based on considering the discrete phase-space as a 2-dimensional vector space labelled by finite fields (for details, see  \cite{GibbonsHW04}). In \cite{Wootters04b} Wootters discusses different geometrical problems associated with this construction.

To define a discrete Wigner function, we need to associate a projector onto a quantum state to each line in discrete phase-space. These will be projectors onto a set of $d+1$ \textit{mutually unbiased bases} (MUB).
Consider two different orthonormal bases $B_1$ and $B_2$:
\begin{eqnarray}
B_1=\{
\left|v_{1,1}\right\rangle,\left|v_{1,2}\right\rangle,...,\left|v_{1,d}\right\rangle
\}, \left|\left\langle
v_{1,i}|v_{1,j}\right\rangle\right|^2 =\delta_{i,j}
,\\
B_2=\{
\left|v_{2,1}\right\rangle,\left|v_{2,2}\right\rangle,...,\left|v_{2,d}\right\rangle
\}, \left|\left\langle
v_{2,i}|v_{2,j}\right\rangle\right|^2
=\delta_{i,j}.\end{eqnarray} These two bases $B_1$ and $B_2$ are \textit{mutually unbiased} if
\begin{equation}
\left|\left\langle
v_{i,j}|v_{k,l}\right\rangle\right|^2=\frac{1}{d} \text{
if } i\neq k.
\end{equation}

Wootters and Fields showed that one can define $(d+1)$ such
mutually unbiased bases for power-of-prime dimension $d$
\cite{WoottersF89}. Mutually unbiased bases have been studied because of their use in a number of quantum information applications, for example quantum cryptography \cite{Bechmann-PasquinucciT00}, quantum state and process tomography \cite{BenderskyPP08} and in the construction of quantum $t$-designs \cite{KlappeneckerR05}, used to estimate averages of functions over quantum states.

To define a discrete Wigner function (DWF) we pick a one-to-one map between the lines in discrete phase-space and the projectors onto a complete set of MUB in the following way:

\begin{itemize}
\item each basis set $B_i$ is associated with one striation $S_i$;
\item each basis vector projector $Q_{i,j} \equiv \left|v_{i,j}\right\rangle\left\langle v_{i,j}\right|$ is associated with a line $\lambda_{i,j}$ (the $j$th line of the $i$th striation).
\end{itemize}
These maps define uniquely the values of the DWF $W_{\beta}$ for all points $\beta$ if we impose the following constraints:
\begin{equation}
Tr\left(
Q_{i,j}
\rho \right)=\sum_{\alpha \in \lambda_{i,j}}W_\alpha\label{defW},
\end{equation}
where $\rho$ is the system's density matrix, and the sum is over phase-space points $\alpha$ in the line $\lambda$ associated with projector $Q_{i,j}$. These requirements amount to demanding that the sum of the Wigner function over each line must be equal to the probability of projecting onto the basis
vector associated with that line.

Note that there are multiple ways of making these associations. In
general, this will lead to different definitions of the DWF using the
same fixed set of MUB. The procedure outlined above then leads
not to a single definition of $W$, but to a class of Wigner
functions instead.

We now define the \emph{phase-space point operator} $A_{\beta}$ associated with phase-space point $\beta$:
\begin{equation}
A_{\beta} \equiv \sum_{\lambda \supset \beta} Q_{\lambda}-I \label{eq defa},
\end{equation}
where the sum is over projectors $Q_{\lambda}$ associated to lines $\lambda$ containing point $\beta$, and $I$ represents identity. The operators $A_{\beta}$ appear naturally when we invert equations (\ref{defW}) to write an expression for $W_{\beta}$ in terms of the MUB projectors:
\begin{equation}
    W_{\beta}=\frac{1}{d}Tr\left(\rho A_{\beta}\right) \label{eq defW},
\end{equation}
We see that the expectation value of $A_{\beta}$ (multiplied by $1/d$) gives the value of the DWF at phase-space point $\beta$.

The $A$ operators form a complete basis in the space of $d\times d$
matrices. It can be shown that the DWF value at point $\alpha$ is simply the expansion coefficient of $\rho$ corresponding to the $A_{\alpha}$ operator:
\begin{equation}
    \rho=\sum_{\alpha}W_{\alpha}A_{\alpha},
\end{equation}
The multiple ways of associating projectors with lines in phase-space result in multiple definitions for $A_{\alpha}$. While a single definition of $W$ requires $d^2$ operators $A_{\alpha}$ (one for each phase-space point), the full set of $A_{\alpha}$ one can define with the same fixed complete set of MUB has $d^{d+1}$ elements. In section \ref{sec qrac} we will make use of the full set of phase-space point operators to obtain a construction for a quantum information application known as quantum random access codes.

\subsection{Negativity and non-classicality}

Cormick \textit{et al.} \cite{CormickGGPP06} have characterized the set of states which have non-negative Wigner functions. These states turn out to have some interesting properties, which we will review here, as they motivate the work reported in the remainder of this paper.

For a $d$-dimensional quantum system, one can find complete sets of $(d+1)$ mutually unbiased bases using the finite-field construction introduced in \cite{GibbonsHW04}. This construction is only valid for power-of-prime $d$, since this is the necessary condition for a finite field to exist. One can then define the set $C_d$ of $d$-dimensional states which have non-negative DWF in all definitions, that is, whose expectation values for \textit{all} phase-space point-operators are non-negative.

In \cite{CormickGGPP06} it was shown that the only pure states in $C_{d}$ are the
MUB projectors, which can always be chosen to be \textit{stabilizer
states}. The stabilizer formalism \cite{Gottesman97} provides a way to represent pure
states in $C_{d}$ using a number of bits which is polynomial in
the number of qubits. Since general pure states require a description which is exponential-size, the set $C_d$ is classical in the sense of having a short description.

For systems of prime dimensions, the two notions of classicality exactly coincide: the only pure states with nonnegative DWF are exactly the stabilizer states. In this context, negativity of any DWF (as witnessed by negativity of one of the $A_{\alpha}$ operators) indicates non-classicality in the sense of the absence of an efficient description using the stabilizer formalism. These results motivated us to investigate the extrema of the discrete Wigner functions.

Before proceeding, we need to review some constructions of complete sets of MUB for $d$-dimensional systems, as these are necessary to define DWF using eqs. (\ref{eq defa}) and (\ref{eq defW}).

\section{Complete sets of mutually unbiased bases}\label{sec mub}

For prime dimension $d$ there is a canonical construction of a complete set of $(d+1)$ MUB's first proposed by Ivanovic \cite{Ivanovic81}. To review this construction, let $\left|v_{r,k}\right\rangle_j$ denote the $j$-th component of the $k$-th vector in the $r$-th basis, $r=0,1,..., d$. The vectors in the complete set of $d+1$ MUB are then:
\begin{eqnarray}
\left|v_{0,k}\right\rangle_j&=&\delta_{jk}\\
\left|v_{1,k}\right\rangle_j&=&\frac{1}{\sqrt{d}}e^{\frac{2\pi i}{d}(j^2+jk)}\\
\vdots\\
\left|v_{r,k}\right\rangle_j&=&\frac{1}{\sqrt{d}}e^{\frac{2\pi i}{d}(rj^2+jk)}\\
\vdots\\
\left|v_{(d-1),k}\right\rangle_j&=&\frac{1}{\sqrt{d}}e^{\frac{2\pi i}{d}((d-1)j^2+jk)}\\
\left|v_{d,k}\right\rangle_j&=&\frac{1}{\sqrt{d}}e^{\frac{2\pi i}{d}jk}
\end{eqnarray}

When the Hilbert space dimension is a prime power, there are different constructions of complete sets of MUB. Let us now review a simple construction of a complete set of MUB for $n$ qubits (Hilbert space dimension $d=2^n$) consisting solely of stabilizer states \cite{LawrenceBZ02}. We start by considering the $4^n$ Pauli operators for $n$ qubits, which are the tensor products of single-qubit Pauli operators $X, Y, Z$ and identity. From this set, remove the identity operator. The remaining $4^n -1$ Pauli operators can be partitioned into $2^n +1$ sets, each containing $2^n-1$ mutually commuting Pauli operators. It was proven in \cite{LawrenceBZ02} that the common eigenstates of the operators in each such set form a basis, and moreover that the $2^n+1$ bases thus defined are mutually unbiased.

We will now provide two examples of this construction, which will be useful to us later on. The first example is a set of $5$ MUB for two qubits, each basis being formed by the common eigeinstates of each row of operators in Table \ref{table mubd4}. In the table, operator $XY$ for example stands for the tensor product of $X$ on the first qubit by $Y$ on the second.

\begin{table}\caption{\label{table mubd4} Set of 5 MUB for two qubits}
\begin{center}
\begin{tabular}{|c|c|c|c|} \hline

1 & $ X  X$ & $ X  1$ & $1  X$ \\ \hline
2 & $ Z  Z$ & $ Z  1$ & $1   Z$\\ \hline
3 & $ Y  Y$ & $ Y  1$ & $1  Y$\\ \hline
4 & $ X  Y$ & $ Y  Z$ & $ Z  X$\\ \hline
5 & $ X  Z$ & $ Y  X$ & $ Z  Y$\\ \hline
\end{tabular}
\end{center}
\end{table}
The second example is a set of 9 MUB for three qubits, comprising the common eigenstates of the operators in each row of Table \ref{table mubd8}.

\begin{table}\caption{\label{table mubd8} Set of 9 MUB for three qubits}
\begin{center}
\begin{tabular}{|c|c|c|c|c|c|c|c|} \hline

1 & $X    X    X $ & $ X    X   1$ & $ X   1   X$ & $ X   1    1$ & $ 1   X    X$ & $ 1   X   1$ & $ 1    1   X$ \\ \hline
2 & $ X    X   Y $ & $ X    Y   X$ & $ Y   X   X$ & $ Y    Y   Y$ & $ Z    Z   1$ & $ Z   1   Z$ & $ 1    Z   Z$ \\ \hline
3 & $ X    X   Z $ & $ X    Y   Y$ & $ Y   Z   1$ & $ Y   1    X$ & $ Z    X   Y$ & $ Z   Y   Z$ & $ 1    Z   X$ \\ \hline
4 & $ X    Y   Z $ & $ X   Z    X$ & $ Y   X   1$ & $ Y    1   Y$ & $ Z    Y   X$ & $ Z   Z   Z$ & $ 1    X   Y$ \\ \hline
5 & $X    Y    1 $ & $ X    1   Z$ & $ Y   X   Y$ & $ Y    Z   X$ & $ Z   X    X$ & $ Z   Z   Y$ & $1    Y    Z$ \\ \hline
6 & $ X    Z   Y $ & $ X    1   Y$ & $ Y   Z   Z$ & $ Y   1    Z$ & $ Z    Z   X$ & $ Z   1   X$ & $ 1    Z   1$ \\ \hline
7 & $ X    Z    Z$ & $ X    Z   1$ & $ Y   Y   Z$ & $ X    Y   1$ & $ Z    X   Z$ & $ Z   X   1$ & $ 1    1   Z$ \\ \hline
8 & $ Y    X   Z $ & $ Y    Y   X$ & $ Y   Z   Y$ & $ Y    1   1$ & $ 1    X   Z$ & $ 1   Y   X$ & $ 1    Z   Y$ \\ \hline
9 & $ Z    Y   Y $ & $ Z    Y   1$ & $ Z   1   Y$ & $ Z   1    1$ & $ 1    Y   Y$ & $ 1   Y   1$ & $ 1    1   Y$ \\ \hline

\end{tabular}
\end{center}
\end{table}

\section{Spectra of phase-space point operators}\label{sec spectra}

As we have seen in the previous sections, the DWF is defined by eq. (\ref{eq defW}) using the phase-space point operators $A$. In this section we calculate the spectra of $A$ for the constructions of complete sets of MUB we reviewed in section \ref{sec mub}. The spectra we tabulate in Table \ref{table spec} agree with the spectra calculated independently by  Appleby \textit{et al.} \cite{ApplebyBC07} for Hilbert space dimensions $d=3,4,5$ and $7$. In Table \ref{table spec} we also report the number of phase-space point operators with each spectrum, an information which will be necessary for the quantum information application described in section \ref{sec qrac}.

We have calculated the full spectra of all $d^{d+1}$ phase-space point operators for $d=2,3,4,5,7$ and $8$, but the latter two cases have too many different spectra for us to reproduce here. We would like to point out only one piece of information about these cases: the extremal eigenvalues found over all the $A_{\alpha}$. For $d=7$ the largest eigenvalue is $\lambda_{max}=2.4178$ and the smallest is $\lambda_{min}=-1$, whereas for $d=8$ the largest is $\lambda_{max}=2.5490$ and the smallest is $\lambda_{min}=-0.9979$.

For the case of two qubits ($d=4$) we calculated the spectra for all six different stabilizer MUB constructions of the kind described in \cite{LawrenceBZ02}, and the spectra found are identical to those of the MUB set in Table \ref{table mubd4}. We have also established that for three qubits ($d=8$), one can find $960$ different stabilizer constructions of this kind. Testing a few of those we found exactly the same spectra as for the example in Table \ref{table mubd8}. Based on this, we conjecture that the phase-space point operator spectra is independent of which MUB set construction one uses.

\begin{table}\caption{\label{table spec} Spectra of phase-space point operators}
\begin{tabular}{|c|c|c|}

\multicolumn{3}{}{}\\
\hline
d & Number & Spectrum \\ \hline
\multirow{1}{*}{$2$} & 8 & $\{\frac{1}{2}+\frac{\sqrt{3}}{2},
\frac{1}{2}-\frac{\sqrt{3}}{2}\}$ \\ \hline
\multirow{2}{*}{$3$} & 9 & $\{-1,1,1\}$  \\
 & 72 & $\{0,\frac{1}{2}+\frac{\sqrt{5}}{2},
\frac{1}{2}-\frac{\sqrt{5}}{2}\}$ \\ \hline

\multirow{3}{*}{$4$} & 320 & $\{-0.50000,  -0.50000,  0.13397,  1.86603\}$ \\
 & 320 & $\{-0.86603,  -0.50000,   0.86603,   1.50000\}$ \\
& 384 &  $\{-0.89680,  -0.14204,  0.27877,   1.76007\}$ \\ \hline

\multirow{9}{*}{$5$} & 1000 & $\{-0.70281, -0.61803, -0.13294, 0.48666, 1.96712\}$ \\
 & 2000 & $\{-0.79859, -0.36221, 0.00000, 0.10661, 2.05419\}$ \\
& 2000 &  $\{-0.83607, -0.81000, 0.00000, 1.05469, 1.59139\}$ \\
 & 3000 &   $\{-0.83726, -0.58152, -0.09576, 0.62870, 1.88584\}$\\
  &1000 &   $\{-0.90039, -0.64018, -0.14531, 1.06785, 1.61803\}$\\
  &3000 &  $\{-0.90932, -0.48701, 0.00000, 0.46853, 1.92780\}$ \\
  &3000 &   $\{-0.94658, -0.51690, -0.18438, 0.93842, 1.70944\}$\\
  &600 &  $\{-1.00000, -0.61803, 0.00000, 1.00000, 1.61803\}$ \\
   &25 &  $\{-1.00000, -1.00000, 1.00000, 1.00000, 1.00000\}$ \\

\hline
\end{tabular}
\end{table}

\section{Extrema of discrete Wigner functions}\label{sec extre}

Unlike probability distributions, we have seen that the discrete Wigner function can assume negative values. What are the extremal values that it can attain? Wootters conjectured that the minimal value $min(W)$ that the discrete Wigner function could assume would be  $-1/d$ for odd-prime Hilbert space dimension $d$ \cite{Wootters05pc}. He also showed that $min(W)=-0.183$ for $d=2$ \cite{Wootters04}, and together with Sussman \cite{WoottersS07} found a maximum value of $max(W)=0.319$ for some particular definitions of discrete Wigner functions for $d=8$. In this section we describe a general method for finding the extrema among all discrete Wigner functions definable with a fixed complete set of MUB, and use it to explicitly calculate the extrema for $d=2,3,4,5,7$ and $8$.

Recall that the phase-space point operator $A_{\alpha}$ associated with the phase-space point $\alpha$ is given as the sum over MUB projectors associated with all phase-space lines that contain $\alpha$:
\begin{equation}
A_{\alpha}=\sum_{\lambda \supset \alpha}Q_{\lambda} - I, \label{defA}
\end{equation}
where the sum is over projectors associated with lines $\lambda$ containing point $\alpha$, and $I$ represents identity.

Given a phase-space point operator $A_{\alpha}$, we want to find the minimum of its expectation value
\begin{equation}
min \left\langle A _{\alpha} \right\rangle = min \left( Tr\left( \sum_{\lambda \supset \alpha}Q_{\lambda} \ketbra{\psi}{\psi}\right)\right)-1. \label{eq trick}
\end{equation}
The minimum results when $\left|{\psi}\right\rangle$ is the eigenvector associated with the smallest eigenvalue $\lambda_{min}$ of $A_{\alpha}$ (see the Appendix for a proof). We can evaluate the spectrum of $A_{\alpha}$ to find its smallest eigenvalue $\lambda_{min}^{\alpha}$. Then, using eq. (\ref{eq defW}),  the most negative value for the discrete WF at point $\alpha$ will be given by
\beq
min(W_{\alpha})=\frac{1}{d} min \left\langle A_{\alpha} \right\rangle = \frac{1}{d}\lambda_{min}^{\alpha}.\label{minla}
\eeq

To find the most negative value for the function, one needs to find all eigenvalues of all possible phase-space point operators. For a $d$-dimensional system there are $d^{d+1}$ different phase-space point operators, only $d^{2}$ of which appear in any single definition of a DWF. Minimizing over $\alpha$, we can use eq. (\ref{minla}) to obtain the smallest value that the function can attain.

The same reasoning can be used to obtain the maxima of the DWF, using the largest eigenvalue of any of the $A_{\alpha}$. Using the spectra tabulated in section \ref{sec spectra} we obtained the extremal values of the DWF for small dimensions $d$, listed in Table \ref{table extr}. The results support Wootters' conjecture for odd-prime $d$.

\begin{table}\caption{\label{table extr}Extremal values for DWF}\begin{center} \begin{tabular} {|c|c|c|} \hline

$d$ & $W_{max}$ & $W_{min}$ \\
\hline \hline
2 & $\frac{1}{4}(1+\sqrt{3}) \simeq0.683$ & $\frac{1}{4}(1-\sqrt{3}) \simeq-0.183$\\ \hline
3 & $\frac{1}{6}(1+\sqrt{5}) \simeq0.539$ & $-\frac{1}{3}$\\ \hline
4 & $0.4665$ & $-0.2242$ \\ \hline
5 & $0.411$ & $-\frac{1}{5}$\\ \hline
7 & $ 0.3454$ & $-\frac{1}{7}$ \\ \hline
8 & $0.3186$ & $-0.1247$ \\ \hline
\end{tabular}
\end{center}
\end{table}

\section{An application: quantum random access codes}\label {sec qrac}

In this section we review the quantum information protocol known as quantum random access codes, and present a code construction that relies on states maximizing the discrete Wigner function. Let us start by recalling what these codes are using a simple example.

Imagine a
situation in which Alice encodes $m$ classical bits into $n$ bits
($m>n$), which she sends to Bob, who will need to know that value of a \textit{single} bit (out of the $m$ possible ones) with a
probability of at least $p$. We may represent such an
encoding/decoding scheme by the notation:
$m\rightarrow n $.

Prior to sending the $n$-bit message,
however, Alice does not know which of the $m$ bits Bob will need to
read out. To maximise the least probability of success $p$, Alice and
Bob need to agree on the use of a particular, efficient $m\rightarrow n$ encoding.

We can consider the
quantum generalization of this situation, in which Alice can send Bob
$n$ qubits of communication, instead of $n$ bits. The idea behind
these so-called \textit{quantum random access codes} (QRACs) is very old by
quantum information standards; it appeared in a paper written
circa 1970 and published in 1983 by Stephen Wiesner
\cite{Wiesner83}.

These codes were re-discovered in
\cite{AmbainisNT-SV99}, where the explicit comparison with classical
codes was made.

\subsection{Example: $3 \to 1$ QRAC with a qubit}
Let us illustrate the idea with a $3 \to 1$ quantum random access code (QRAC) that encodes three bits into a single qubit. This QRAC was attributed to Isaac Chuang in ref. \cite{AmbainisNT-SV99}.

\begin{figure}[b]
{\includegraphics[scale=0.5]{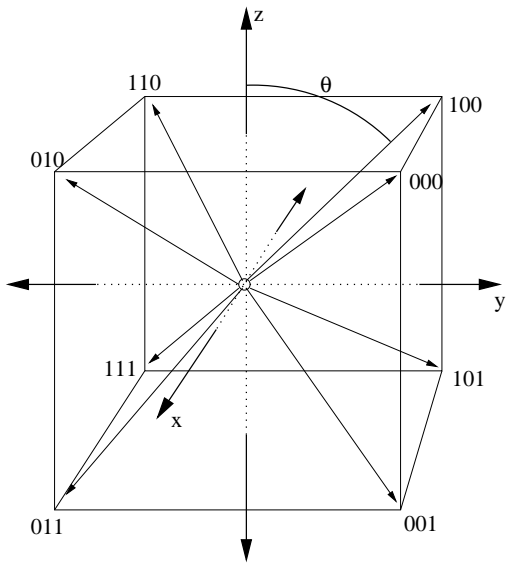}}
\caption[Encoding states for the qubit $3 \to 1$ QRAC. ]{\label{fig qcca}Encoding states for the $3 \to 1$ QRAC using a single
qubit. Alice prepares one out of eight states on the vertices of a
cube inscribed within the Bloch sphere, depending on her three-bit
string. The angle $\theta$ is such that
$\cos^2(\theta/2)=1/2+\sqrt(3)/6 \simeq 0.79$, which is the probability
of Bob correctly decoding a single bit out of the three.}
\end{figure}

Instead of concerning ourselves with the least decoding probability of success $p$, we will use as figure of merit the average probability of success $p_q$. With three bits, Alice has $2^3=8$ possible bit-strings $b_0b_1b_2$. For each
possibility she will prepare one particular state from the set
depicted in Figure \ref{fig qcca}. These states lie on the vertices of a
cube inscribed within the Bloch sphere, which is the representation of one-qubit pure states using spherical-coordinate angles $\theta, \phi$:
\begin{equation}
\left|\psi(\theta,\phi)\right\rangle=\cos(\theta/2)\left|0\right\rangle+\exp(i\phi)\sin(\theta/2)\left|1\right\rangle .
\end{equation}

If Bob wants to read out bit $b_0$ he measures along the $x$-axis and associates a positive result
with $b_0=0$. To read bits $b_1$ [$b_2$] he measures along the
$y$-axis [$z$-axis] and again associates a positive result with
$b_1=0$ [$b_2=0$]. It is easy to see that Bob's average probability of success is
given by $p_q=\cos^2(\theta/2)=1/2+\sqrt(3)/6 \simeq 0.79$, where the
angle $\theta$ is given in the caption to Figure \ref{fig qcca}. The optimal classical $3 \to 1$
random access code succeeds only with average probability $p_c=0.75$, as can
be checked easily through a search over all deterministic protocols.

Note that the QRAC just presented uses decoding measurements which are projections onto the canonical set of MUB for a qubit, that is, the $X, Y$ and $Z$ bases. The encoding states are found by optimizing the probability of correctly decoding each coding state.

\subsection{A QRAC construction from $A_{\alpha}$}

The full set of $d^{d+1}$ phase-space point operators $A_{\alpha}$ can be used to build a particularly symmetric set of quantum random access codes. The encoding states will be those maximizing each $\left\langle A_{\alpha} \right\rangle$. Our results from section \ref{sec extre} showed that those are the largest-eigenvalue eigenstates of each $A_{\alpha}$.

Our goal is to construct a QRAC that encodes $(d+1)$ messages with $d$ possible values each, using a single quantum $d$-level system that will be sent by Alice and measured by Bob. As in the case with  qubits, our strategy will be for Bob to decode by performing projective measurements onto one out of the $(d+1)$ MUB that exist for this $d$-dimensional system (for power-of-prime $d$).  Alice has to find $d^{d+1}$ different encoding states, each of which will decode correctly with the highest possible probability.

As we have shown in section \ref{sec extre}, the pure quantum state $\left|\psi_{\alpha}\right\rangle$ that maximizes $\left\langle A_{\alpha} \right\rangle$ is the eigenvector with largest eigenvalue $\lambda_{max}^{\alpha}$ of $\left\langle A_{\alpha} \right\rangle$. Bob's decoding procedure involves measuring the encoding state $\left|\psi_{\alpha}\right\rangle$ onto one of the $(d+1)$ MUB's. His average probability of successfully decoding will be the average $\frac{1}{d+1} \sum_{\lambda \supset \alpha} \langle Q_{\lambda} \rangle$. To maximize this decoding probability for encoding state $\left|\psi_{\alpha}\right\rangle$, we need to maximize
\begin{eqnarray}
max\left(\frac{1}{d+1}\sum_{\lambda \supset \alpha} \langle Q_{\lambda} \rangle\right)&=& \frac{1}{d+1}max (\left\langle A_{\alpha} \right\rangle+1) \\
&=&\frac{1}{d+1}(\lambda_{max}^{\alpha}+1).
\end{eqnarray}
The average performance of this QRAC protocol can be found by averaging the probability of success of the $d^{d+1}$ encoding states, each corresponding to one phase-space point operator $A_{\alpha}$:
\beq
p_q =\frac{1}{d^{d+1}} \frac{1}{d+1}\sum_{\alpha} (\lambda_{max}^{\alpha}+1). \label{eq pq}
\eeq
We see that the protocol's average success rate $p_q$ depends on the value of the sum of the largest eigenvalues of each of the $d^{d+1}$ possible phase-space point operators $A_{\alpha}$.

Using the MUB constructions described in section \ref{sec mub} and the point operator spectra calculated in section \ref{sec spectra}, we were able to compute $p_q$ for a $(d+1)\to d$ QRAC using systems of dimension $d=2,3,4,5,7$ and $8$. The results are summarized in Table \ref{table pqd}. The construction recovers the known success rate of the $3 \to 1$ QRAC with a qubit.

\begin{table}\caption{\label{table pqd} QRAC success rate $p_q$}
\begin{center}
\begin{tabular} {|c|c|} \hline
Dimension & $p_q$ \\
\hline \hline
$d=2$ &  $\frac{1}{6}(3+\sqrt{3})\simeq0.789$ \\ \hline
$d=3$ & $\frac{7}{18}+\frac{\sqrt{5}}{2}\simeq0.637$ \\ \hline
$d=4$ &  $0.5424$ \\ \hline
$d=5$ & $0.4700$ \\ \hline
$d=7$ &  $0.3720$ \\ \hline
$d=8$ & $0.3372$\\ \hline
\end{tabular}
\end{center}
\end{table}

It is clear that this construction can be extended to higher-dimensional systems, provided a DWF can be defined for them. This will be the case for power-of-prime dimensions $d$, using for example the finite-field construction given in \cite{GibbonsHW04}.

It would be interesting to investigate how these protocols fare against the optimal classical protocols. This would require evaluating the optimal probability of success for a $(d+1)\rightarrow d$ $d$-level classical random access code, something that to our knowledge has not been done for $d>2$. In \cite{AmbainisNT-SV99} some asymptotic results were obtained for large $d$, indicating that while there may be an advantage of quantum over classical for small dimensions $d$, this advantage practically disappears in the asymptotic regime.

\section{Conclusion}

We have reviewed the definition of discrete Wigner functions given in \cite{GibbonsHW04}, used to describe quantum systems in a discrete phase-space. We calculated the spectra of phase-space point operators for small dimensions, and used them to obtain the extrema of the discrete Wigner function for systems in Hilbert-space dimensions $d=2,3,4,5,7$ and $8$. We then described the protocol known as quantum random access codes (QRAC), and used the phase-space point operators to find encoding states and to obtain the efficiency of a QRAC construction whose encoding states maximize the discrete Wigner function at each phase-space point.
\bigskip

\appendix*
\section{}

Let $\rho$ be a density matrix, hence positive semi-definite, Hermitian and with unit trace. Let $\left|\psi\right\rangle$ be a pure state. In this Appendix we prove that the extrema of
\begin{equation}
\left\langle \phi | \rho\right| \phi\rangle \label{tomin}
\end{equation}
are obtained when $\left| \phi\right\rangle$ is an eigenstate corresponding to extremal eigenvalues of $\rho$.

%Let $\lambda$ be the (non-negative) eigenvalues of $\rho$. Let us expand a general pure state $\left|\psi\right\rangle$ in the basis that diagonalizes $\rho$:
%\begin{equation}
%\left|\psi\right\rangle = \sum_i c_i \left| i \right\rangle.
%\end{equation}
%We can evaluate expression (\ref{tomin}) in the same basis, obtaining
%\begin{equation}
%\left\langle \psi | \rho\right| \psi\rangle = \sum_i \lambda_i \left| c_i\right|^2.
%\end{equation}
We start by showing that the state that maximizes this expression is the eigenstate $\left|\lambda_{max}\right\rangle$ corresponding to the largest eigenvalue $\lambda_{max}$ of $\rho$. It is easy to evaluate eq. (\ref{tomin}) for $\left|\phi\right\rangle = \left|\lambda_{max}\right\rangle$:
\begin{equation}
\left\langle \lambda_{max} | \rho\right| \lambda_{max}\rangle = \lambda_{max} \left\langle \lambda_{max}|\lambda_{max} \right \rangle = \lambda_{max}.
\end{equation}
Let us now consider the expansion of a general state $\left|\phi\right\rangle=\sum_i d_i \left| i\right\rangle$ in the basis that diagonalizes $\rho$. For $\left|\phi\right\rangle$, the expression we are trying to maximize takes the value:
\begin{equation}
\left\langle \phi | \rho\right| \phi\rangle = \sum_i \lambda_i \left| d_i\right|^2= \lambda_{max}(1-\Delta)+\sum_{j\neq \lambda_{max}}\lambda_i \left| d_i\right|^2 ,
\end{equation}
where $\lambda_i$ are the eigenvalues of $\rho$. We have rewritten the sum to single out the term corresponding to $\lambda_{max}$, and defined $\Delta>0 $ such that $(1-\Delta)=|d_{max}|^2$. Now, because all $\lambda_i \le \lambda_{max}$ (by definition of $\lambda_{max}$), we have
\begin{equation}
\left\langle \phi | \rho\right| \phi\rangle =  \lambda_{max}(1-\Delta)+\sum_{j\neq \lambda_{max}}\lambda_i \left| d_i\right|^2 \le \lambda_{max}.
\end{equation}
So we have proven that any set of coefficients different from those of $\left| \lambda_{max}\right\rangle$ leads to a smaller expectation value for $\rho$, and so $\left| \lambda_{max}\right\rangle$ maximizes this expectation value. A similar argument can be made to prove that the state that minimizes $\left\langle \phi | \rho\right| \phi\rangle$ is the eigenstate corresponding to the smallest eigenvalue of $\rho$.

The claim following eq. (\ref{eq trick}) is justified by applying the result above to $\rho=\frac{1}{d+1}\sum_{\lambda \supset \alpha}{Q_{\lambda}}$.

\begin{acknowledgments}
The authors are grateful to Ilaria Cardinali and Enrico Martinelli for helpful discussions. E.F.G. acknowledges support from Brazilian funding agencies FAPERJ and CNPq. S.S. acknowledges financial support from DTO-ARO, ORDCF, Ontario-MRI, CFI, CIFAR, and MITACS.
\end{acknowledgments}

\bibliographystyle{unsrt}
%\bibliography{wigner}

\end{document}